\documentclass[entropy,preprints,accept,moreauthors,pdftex,10pt,a4paper]{mdpi}
\firstpage{1} 
\articlenumber{}
\doinum{}
\pubvolume{}
\pubyear{}
\copyrightyear{}
\externaleditor{}
\history{}

\pdfoutput=1

\usepackage{amssymb}
\usepackage{theoremref}
\usepackage{relsize}
\usepackage{subfloat}
\usepackage{subcaption}
\usepackage[all]{hypcap}

\renewcommand{\textcolor}[2]{#2}

\Title{Comparing Information-Theoretic Measures of Complexity in Boltzmann Machines}

\Author{Maxinder S. Kanwal $^{1,}$*, Joshua A. Grochow $^{2,3}$ and Nihat Ay $^{3,4,5}$}
\AuthorNames{Maxinder S. Kanwal, Joshua A. Grochow and Nihat Ay}

\address{%
$^{1}$ \quad University of California, Berkeley\\
$^{2}$ \quad University of Colorado, Boulder\\
$^{3}$ \quad Santa Fe Institute\\
$^{4}$ \quad Max Planck Institute for Mathematics in the Sciences\\
$^{5}$ \quad University of Leipzig
}

\corres{Correspondence: mkanwal@berkeley.edu}

\abstract{
In the past three decades, many theoretical measures of complexity have been proposed to help understand complex systems.  In this work, for the first time, we place these measures on a level playing field, to explore the qualitative similarities and differences between them, \textcolor{red}{and their shortcomings}.  Specifically, using the Boltzmann machine architecture (a fully connected recurrent neural network) with uniformly distributed weights as our model of study, we \textcolor{red}{numerically} measure how complexity changes as a function of network dynamics and network parameters.  We apply an extension of one such information-theoretic measure of complexity to understand incremental Hebbian learning in Hopfield networks, a fully recurrent architecture model of autoassociative memory.  In the course of Hebbian learning, the total information flow reflects a natural upward trend in complexity as the network attempts to learn more and more patterns.
}


\begin{document}

%%%%%%%%%%%%%%%%%%%%%%%%%%%%%%%%%%%%%%%%%%

\section{Introduction}

Many systems, across a wide array of disciplines, have been labeled ``complex''.  The striking analogies between these systems \cite{complex_sys1, complex_sys2} beg the question: What collective properties do complex systems share and what quantitative techniques can we use to analyze these systems as a whole?  With new measurement techniques and ever-increasing amounts of data becoming available about larger and larger systems, we are in a better position than ever before to understand the underlying dynamics and properties of these systems.

While few researchers agree on a specific definition of a complex system, common terms used to describe complex systems include ``emergence'' and ``self-organization'', which characterize high-level properties in a system composed of many simpler sub-units.  Often these sub-units follow local rules that can be described with much better accuracy than those governing the global system.  Most definitions of complex systems include, in one way or another, the hallmark feature that the whole is more than the sum of its parts.

In the unified study of complex systems, a vast number of measures have been introduced to concretely quantify an intuitive notion of complexity (see, e.g.,  \cite{lloyd, shalizi}).  As Shalizi points out \cite{shalizi}, among the plethora of complexity measures proposed, roughly, there are two main threads: those that build on the notion of Kolmogorov complexity and those that use the tools of Shannon's information theory.  There are many systems for which the nature of their complexity seems to stem either from logical/computational/descriptive forms of complexity (hence, Kolmogorov complexity) and/or from information-theoretic forms of complexity.  In this paper we focus on information-theoretic measures.

While the unified study of complex systems is the ultimate goal, due to the broad nature of the field, there are still many sub-fields within complexity science \cite{complex_sys1, complex_sys2, crutchfield}.  One such sub-field is the study of networks, and in particular, stochastic networks (broadly defined).  Complexity in a stochastic network is often considered to be directly proportional to the level of stochastic interaction of the units that compose the network---this is where tools from information theory come in handy.

\subsection{Information-Theoretic Measures of Complexity} \label{intro}
Within the framework of considering stochastic interaction as a proxy for complexity, a few candidate measures of complexity have been developed and refined over the past decade.  There is no consensus best measure, as each individual measure frequently captures some aspects of stochastic interaction better than others.

\textls[-10]{In this paper, we empirically examine four measures (described in detail later): (1)~multi-information,} (2)~synergistic information, (3)~total information flow, (4)~geometric integrated information.  Additional \textcolor{red}{notable} information-theoretic measures that we do not examine include those of Tononi et al., first proposed in \cite{tononi_original} and most recently refined in \cite{tononi}, as a measure of consciousness, as well as similar measures of integrated information described by Barrett \& Seth \cite{Seth}, and Oizumi~\textit{et~al.}~\cite{amari_original}.

The term ``humpology,'' first coined by Crutchfield \cite{crutchfield}, attempts to qualitatively describe a long and generally understood feature that a natural measure of complexity ought to have.  In particular, as stochasticity varies from $0\%$ to $100\%$, the structural complexity should be unimodal, with a maximum somewhere in between the extremes \cite{complex_sys3}.  For a physical analogy, consider the spectrum of molecular randomness spanning from a rigid crystal \textcolor{red}{(complete order)} to a random gas \textcolor{red}{(complete disorder)}.  At both extremes, we intuitively expect no complexity: a crystal has no fluctuations, while a totally random gas has complete unpredictability across time.  Somewhere in between, \textcolor{red}{structural} complexity will be maximized (assuming it is always finite).

We now describe the four complexity measures of interest in this study.  
\textcolor{red}{We assume a compositional structure of the system and consider a finite set 
$V$ of nodes. With each node $v \in V$, we associate a finite set ${\Bbb X}_v$ of states. In the prime example of this article, the Boltzmann machine, we  
have $V = \{1,\dots,N\}$, and $\mathsf{\Bbb X}_v = \{\pm 1\}$ for all $v$. For any subset $A \subseteq V$, we define the state set of all nodes in $A$ as 
the Cartesian product ${\Bbb X}_A := \prod_{v \in A} {\Bbb X}_v$ and use the abbreviation ${\Bbb X} := {\Bbb X}_V$. In what follows, we want to consider stochastic processes in 
${\Bbb X}$ and assign various complexity measures to these processes.   
With a probability vector $p(x)$, $x \in {\Bbb X}$, and a
stochastic matrix $P(x , x')$, $x,x' \in {\Bbb X}$, we associate a pair $(X,X')$ of random variables satisfying
\begin{equation} \label{pair}
     p(x,x') := {\operatorname{Pr}}(X = x , X' = x') = p(x) P(x, x'), \qquad x,x' \in {\Bbb X}.
\end{equation}
Obviously, any such pair of random variables satisfies ${\operatorname{Pr}}(X= x) = p(x)$, and ${\operatorname{Pr}}(X' = x' | X = x) = P(x,x')$ whenever $p(x) > 0$. As we
want to assign complexity measures to transitions of the system state in time, we also use the more suggestive notation $X \to X'$ instead of $(X,X')$. If we iterate the transition, we obtain a Markov chain $X_n = {(X_{n,v})}_{v \in V}$, $n = 1,2,\dots$, in ${\Bbb X}$, with
\begin{equation} \label{chain}
    p(x_1,x_2,\dots,x_n) := {\operatorname{Pr}}(X_1 = x_1, X_2 = x_2, \dots, X_n = x_n) =  p(x_1) \prod_{k = 2}^{n} P(x_{k-1} , x_{k}), \qquad n = 1,2,\dots,  
\end{equation}
where, by the usual convention, the product on the right-hand side of this equation equals one if the index set is empty, that is, for $n = 1$. Obviously, we have 
${\operatorname{Pr}}(X_1= x) = p(x)$, and ${\operatorname{Pr}}(X_{n + 1} = x' | X_n = x) = P(x,x')$ whenever $p(x) > 0$.  
Throughout the whole paper, we will assume that the probability vector $p$ is stationary with respect to the stochastic matrix $P$. More precisely, we assume 
that for all $x' \in {\Bbb X}$ the following equality holds:
\[
       p(x') = \sum_{x \in {\Bbb X}} p(x)  P(x , x'). 
\]
With this assumption, we have ${\operatorname{Pr}}(X_n = x) = p(x)$, and the distribution of $(X_n, X_{n + 1})$ does not depend on $n$. This will allow us to restrict attention to only one transition 
$X \to X'$. In what follows, we define various information-theoretic measures associated with such a transition.     
}

\subsubsection{Multi-Information, $MI$}
The multi-information is a measure proposed by McGill \cite{mcgill} that captures the extent to which the whole is greater than the sum of its parts when averaging over time.  For the above random variable $X$, it is defined as
\begin{equation} \label{multiinf}
MI(X) \triangleq \sum\limits_{v \in V} H(X_v) - H(X),
\end{equation}
\textcolor{red}{where the Shannon entropy $H(X) = -\sum_{x \in \Bbb{X}} p(x) \log{p(x)}$.  (Here, and throughout this article, we take logarithms with respect to base $2$.)}  It holds that $MI(X)=0$ if and only if all of the parts, $X_i$, are mutually independent.

\subsubsection{Synergistic Information, $SI$}
The synergistic information, proposed by Edlund et al. \cite{adami}, measures the extent to which the \textcolor{red}{\emph{(one-step) predictive information} of the whole is greater than that of the parts.  (For details related to the predictive information, see \cite{bialek, grassberger, crutchfield2003}.)}  It builds on the multi-information by including the dynamics through time in the measure:
\begin{equation}
SI(X \rightarrow X^\prime)\triangleq I(X;X^\prime) - \sum\limits_{v \in V} I(X_v;X^{\prime}_v),
\end{equation}
where $I(X;X^\prime)$ denotes the mutual information between $X$ and $X^\prime$.  One potential issue with the synergistic information is that it may be negative.  This is not ideal, as it is difficult to interpret a negative value of complexity.  Furthermore, a preferred baseline minimum value of $0$ serves as a reference point against which one can objectively compare systems.

The subsequent two measures (total information flow and geometric integrated information) have geometric formulations that make use of tools from information geometry.  In information geometry, the Kullback-Leibler divergence (KL divergence) is used to measure the dissimilarity between two discrete probability distributions. 
\textcolor{red}{Applied to our context, we measure the dissimilarity between two stochastic matrices $P$ and $Q$ with respect to $p$ by 
\begin{equation} \label{KL}
D_{KL}^p(P \| Q) =  \sum\limits_{x \in {\Bbb X}} p(x) \sum_{x' \in {\Bbb X}} P(x , x') \log {\frac{P(x , x')}{Q(x , x')}}.
\end{equation}
For simplicity, let us assume that $P$ and $Q$ are strictly positive and that $p$ is the stationary distribution of $P$. In that case, we do not explicitly refer to the stationary distribution $p$ and simply write $D_{KL}(P \| Q)$. The KL divergence between $P$ and $Q$ can be interpreted by considering their corresponding Markov chains with distributions (\ref{chain}) (e.g., see \cite{nagaoka} for additional details on this formulation). Denoting the chain of $P$ by $X_n$, $n = 1,2, \dots$, and the chain of $Q$ by $Y_n$, $n = 1,2, \dots$, with some initial distributions $p_1$ and $q_1$, respectively, we obtain
\begin{eqnarray*}
   \lefteqn{\frac{1}{n} \sum_{x_1,x_2,\dots,x_n} {\operatorname{Pr}}(X_1 = x_1, X_2 = x_2,\dots,X_n = x_n) 
                    \log \frac{{\operatorname{Pr}}(X_1 = x_1, X_ 2 = x_2,\dots,X_n = x_n)}{{\operatorname{Pr}}(Y_1 = x_1, Y_2 = x_2,\dots, Y_n = x_n)}} \\
    & = & \frac{1}{n} \left( \sum_{x_1} {\operatorname{Pr}}(X_1 = x_1) \log \frac{{\operatorname{Pr}}(X_1 = x_1)}{{\operatorname{Pr}}(Y_1 = x_1)} + \right. \\
    &    &  \qquad\qquad \left. \sum_{k = 1}^{n-1} \sum_{x} {\operatorname{Pr}}(X_k = x) \sum_{x'} {\operatorname{Pr}}(X_{k+1} = x' | X_{k} = x) 
                           \log \frac{{\operatorname{Pr}}(X_{k+1} = x' | X_{k} = x)}{{\operatorname{Pr}}(Y_{k+1} = x' | Y_{k} = x)}\right) \\
    & = & \frac{1}{n} \sum_{x} p_1(x) \log \frac{p_1(x)}{q_1(x)}   +  \frac{n - 1}{n} \sum_{x} p(x) \sum_{x'} P(x,x') \log \frac{P(x,x')}{Q(x,x')} \\
    & \stackrel{n \to \infty}{\to} & D_{KL}(P \| Q). 
\end{eqnarray*}
We can use the KL divergence  (\ref{KL}) to answer our original question---\textit{To what extent is the whole greater than the sum of its parts?}---by comparing a system of interest to its most similar (least dissimilar) system whose whole is exactly \textit{equal} to the sum of its parts.  When comparing a transition $P$ to $Q$ using the KL divergence, one measures the amount of information lost when $Q$ is used to approximate $P$.  Hence, by constraining $Q$ to be equal to the sum of its parts, we can then arrive at a natural measure of complexity by taking the \textit{minimum} extent to which our distribution $P$ is greater (in the sense that it contains more information) than some distribution $Q$, since $Q$ represents a system of zero complexity.  Formally, one defines a manifold $\mathcal{S}$ of so-called ``split'' systems consisting of all those distributions that are equal to the sum of their parts, and then measures the minimum distance to that manifold:
\begin{equation}\label{complexity}
Complexity(P)\triangleq \min_{Q \in \mathcal{S}} D_{KL}(P \| Q).
\end{equation}

It is important to note here that there are many different viable choices of split manifold $\mathcal{S}$.  This approach was first introduced by Ay for a general class of manifolds ${\mathcal S}$ \cite{ay1}. Amari~\cite{info_geom} and Oizumi~et~al.~\cite{amari} proposed variants of this quantity as measures of information integration. In what follows, we consider measures of the form \eqref{complexity} for two different choices of $\mathcal{S}$.}

\enlargethispage*{1.8\baselineskip}

\subsubsection{Total Information Flow, $IF$}\label{total info flow}
The total information flow, also known as the stochastic interaction, expands on the multi-information (like $SI$) to include temporal dynamics.  Proposed by Ay in \cite{ay1, ay2}, the measure can be expressed by constraining \textcolor{red}{{$Q$} to the manifold of distributions, $\mathcal{S}^{(1)}$, where there exists functions $f_v(x_v, x_v')$, $v \in V$, such that {$Q$} is of the form:
\begin{equation}\label{ay_constraint}
Q(x,x') = Q((x_v)_{v \in V} , (x_v')_{v \in V}) = \frac{e^{\sum_{v \in V} f_v(x_v,x_v')}}{Z(x)},
\end{equation}
where $Z(x)$ denotes the partition function that properly normalizes the distribution.  Note that any stochastic matrix of this kind satisfies the property that $Q(x,x') = \prod_{v \in V} \operatorname{Pr}(X'_v = x'_v \mid X_v = x_v)$.}  This results in
\begin{align}
IF(X \rightarrow X^\prime)& \triangleq \min_{\textcolor{red}{Q \in \mathcal{S}^{(1)}}} D_{KL}( \textcolor{red}{P} \| \textcolor{red}{Q})\\
&= \sum\limits_{v \in V} H(X^{\prime}_v \mid X_v) - H(X^\prime \mid X). \label{entropicrep}
\end{align}

\begin{figure}[p]
\centering
\includegraphics[width=0.8\textwidth]{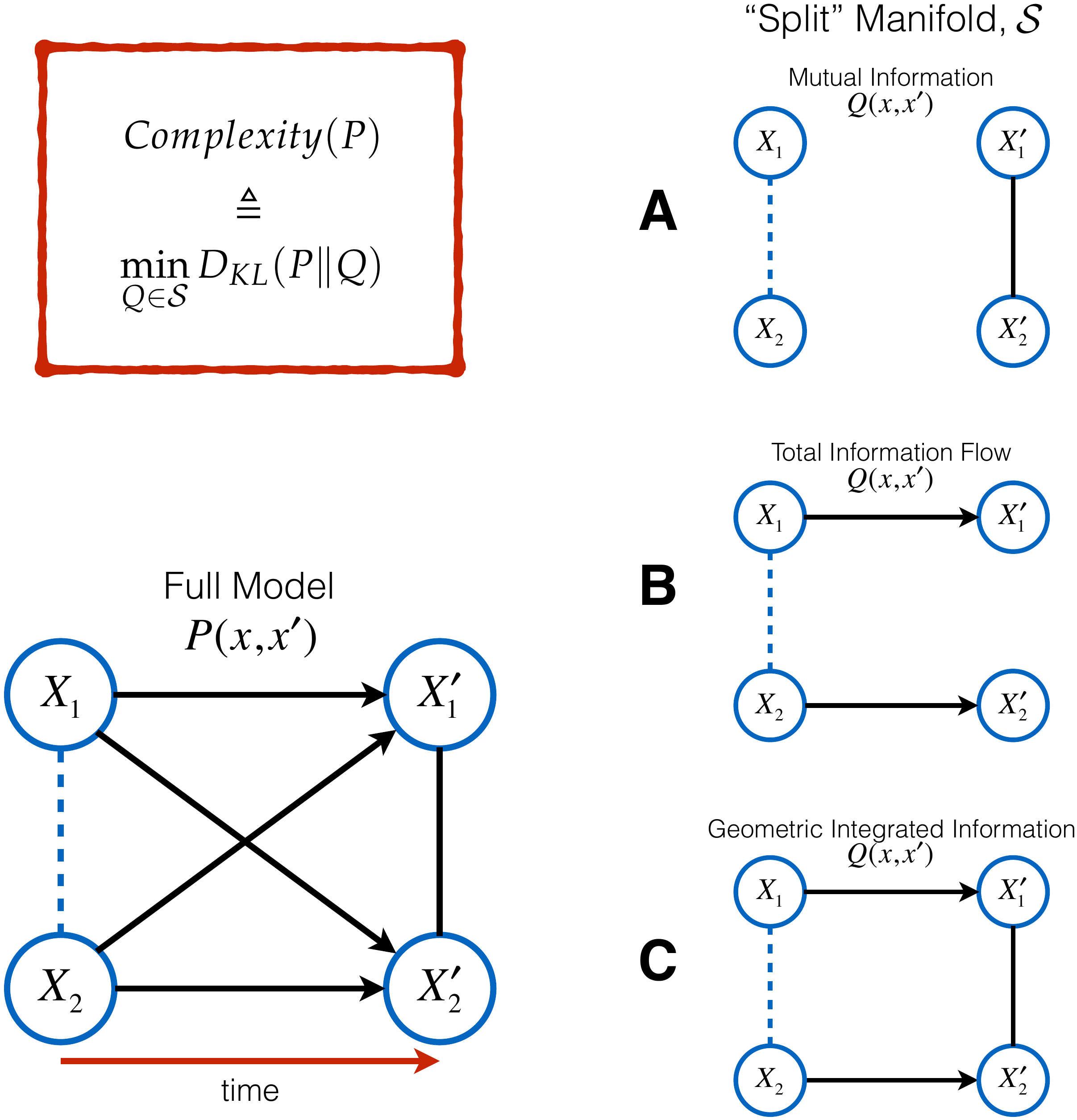}
\caption{Using graphical models, we can visualize different ways to define the ``split'' constraint on manifold $\mathcal{S}$ in \eqref{complexity}.  Here we consider a two-node network $X=(X_1, X_2)$ and its spatio-temporal stochastic interactions.  \textbf{(A)}~$I(X;X^\prime)$ uses constraint \eqref{I_constraint}.  \textbf{(B)}~$IF(X \rightarrow X^\prime)$ uses constraint \eqref{ay_constraint}.  \textbf{(C)}~$\Phi_G(X \rightarrow X^\prime)$ uses constraint \eqref{amari_constraint}.  \textcolor{red}{Dashed lines represent correlations that either may or may not be present in the input distribution $p$.  We do not represent these correlations with solid lines in order to highlight (with solid lines) the structure imposed on the stochastic matrices.}  Adapted and modified from \cite{amari}.}
\label{GraphicalModel}
\end{figure}

The total information flow is non-negative, as are all measures that can be expressed as a KL~divergence.  One issue of note, as pointed out in \cite{info_geom, amari}, is that 
$IF(X \rightarrow X^\prime)$ can exceed $I(X;X^\prime)$.  \textcolor{red}{One can formulate the mutual information $I(X;X')$ as}
\begin{equation}
I(X;X^\prime) = \min_{\textcolor{red}{Q \in \mathcal{S}^{(2)}}} D_{KL}( \textcolor{red}{P}  \| \textcolor{red}{Q}),
\end{equation}
where \textcolor{red}{${\mathcal S}^{(2)}$ consists of stochastic matrices $Q$ that satisfy 
\begin{equation}\label{I_constraint}
Q(x,x') = Q((x_v)_{v \in V} , (x_v')_{v \in V}) = 
\frac{e^{f_V(x')}}{Z(x)},
\end{equation}
for some function $f_V(x')$.  Under this constraint, $Q(x,x^\prime) = \operatorname{Pr}(X^\prime = x^\prime)$.  In other words, all spatio-temporal interactions $X \rightarrow X^\prime$ are lost.} Thus, it has been postulated that no measure of \textcolor{red}{information integration, such as the total information flow,} 
should exceed the mutual information \cite{amari_original}.  The cause of this violation in the total information flow is due to the fact that $IF(X \rightarrow X^\prime)$ quantifies same-time interactions in $X^\prime$ \textcolor{red}{(due to the lack of an undirected edge in the output in Figure \ref{GraphicalModel}B)}. \textcolor{red}{Consider, for instance, a stochastic matrix $P$ that satisfies (\ref{I_constraint}), $P(x,x') = p(x')$ for some probability vector $p$. In that case we have $I(X;X') = 0$. Yet, \eqref{entropicrep} then reduces to the multi-information \eqref{multiinf} of $X' = (X_v')_{v \in V}$, which is a measure of stochastic dependence.}

\subsubsection{Geometric Integrated Information, $\Phi_G$}
\textcolor{red}{In order to obtain a measure of information integration that does not exceed the mutual information $I(X;X')$,}
Amari \cite{info_geom} (Section~6.9) defines $\Phi_G(X \to X')$ as \textcolor{red}{
\begin{equation}
\Phi_G(X \to X') \triangleq \min_{Q \in \mathcal{S}^{(3)}} D_{KL}( P \| Q),
\end{equation}
where $\mathcal{S}^{(3)}$ contains not only the split matrices \eqref{ay_constraint}, but also those matrices that satisfy \eqref{I_constraint}. More precisely,
the set $\mathcal{S}^{(3)}$ consists of all stochastic matrices for which there exists functions $f_v(x_v, x_v')$, $v \in V$, and $f_V(x')$ such that
\begin{equation}\label{amari_constraint}
Q(x,x') = Q((x_v)_{v \in V} , (x_v')_{v \in V}) = 
\frac{e^{\sum_{v \in V} f_v(x_v,x_v') + f_V(x')}}{Z(x)}.
\end{equation}
}
Here, \textcolor{red}{$Q$ belongs to} the set of \textcolor{red}{matrices} where only time-lagged interactions are removed.  
\textcolor{red}{Note that the manifold ${\mathcal S}^{(3)}$ contains ${\mathcal S}^{(1)}$, the model of split matrices used for $IF$, as well as ${\mathcal S}^{(2)}$, the manifold used for the mutual information.}  
This measure thus satisfies both postulates that $SI$ and $IF$ only partially satisfy:
\begin{equation}\label{postulate}
0 \leq \Phi_G(X \to X') \leq I(X;X^\prime).
\end{equation}
However, unlike $IF(X \to X')$, there is no closed-form expression to use when computing $\Phi_G(X~\to~X')$.  In this paper, we use the iterative scaling algorithm described in \cite{IS} (Section~5.1) to compute $\Phi_G(X \to X')$ for the first time in concrete systems of interest.

\textcolor{red}{Note that, in defining $\Phi_G(X \to X')$, the notion of a split model used by Amari \cite{info_geom} is related, but not identical, to that used by Oizumi et al. \cite{amari}.  The manifold considered in the latter work is defined in terms of conditional independence statements and forms a curved exponential family.}

\textcolor{red}{In the remainder of this article, we also use the shorthand notation $MI$, $SI$, $IF$, and $\Phi_G$, without explicit reference to $X$ and $X'$, as already indicated in each measure's respective subsection heading.  We also use $I$ as shorthand for the mutual information.}

\subsection{Boltzmann Machine} \label{rnn}
In this paper, we look at the aforementioned candidate measures in a concrete system in order to gain an intuitive sense of what is frequently discussed at a heavily theoretical and abstract level.  Our system of interest is the Boltzmann machine (a fully-recurrent neural network with sigmoidal activation units).

We parameterize a network of $N$ binary nodes by $W \in \mathbb{R}^{N \times N}$, which denotes the connectivity matrix of weights between each directed pair of nodes.  Each node $i$ takes a value $X_i \in \{\pm 1\}$, and updates to $X_i' \in \{\pm 1\}$ according to:
\begin{equation}
\operatorname{Pr}(X^{\prime}_i = +1 \mid X) = sigmoid\bigg(-2\beta \sum\limits_{j=1}^N w_{ji}\cdot X_j\bigg),
\end{equation}
where
$sigmoid(t) = \frac{1}{1+e^{-t}}$,
$\beta$ denotes a global inverse-temperature parameter, and
$w_{ji}$ denotes the directed weight from $X_j$ to $X_i$.

\newpage

\begin{figure}[!ht]
\centering
\includegraphics[width=0.55\textwidth]{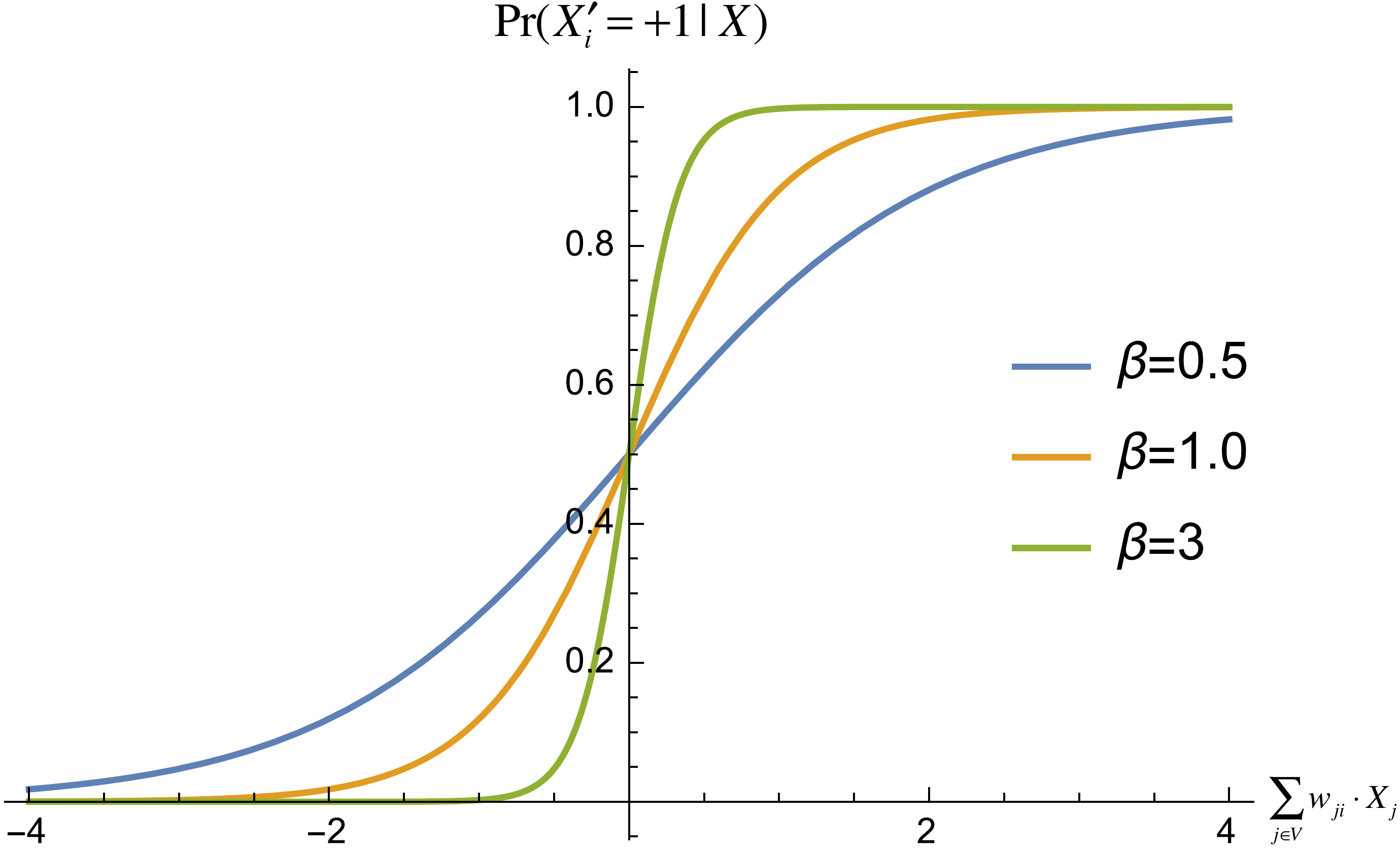}
\caption{The sigmoidal update rule as a function of the inverse-global temperature: As $\beta$ increases, the stochastic update rule becomes closer to the deterministic one given by a step function.}
\label{sigmoid}
\end{figure}

This stochastic update rule implies that every node updates probabilistically according to a weighted sum of the node's parents (or inputs), which, in the case of our fully recurrent neural network, is every node in the network.  Every node $i$ has some weight, $w_{ij}$, with which it influences node $j$ on the next update.  As the weighted sum of the inputs to a node becomes more positive, the likelihood of that node updating to the state $+1$ increases.  The opposite holds true as the weighted sum becomes more negative, as seen in Figure \ref{sigmoid}.  The weights between nodes are a set of parameters we are free to tune in the network.

The second tunable parameter in our network is $\beta$, commonly known as the global inverse-temperature of the network.  $\beta$ effectively controls the extent to which the system is influenced by random noise: It quantifies the system's deviation from deterministic updating.  In networks, the noise level directly correlates with what we call the ``pseudo-temperature'' $T$ of the network, where $T = \frac{1}{\beta}$.  To contextualize what $T$ might represent in a real-life complex system, consider the example of a biological neural network, where we can think of the pseudo-temperature as a parameter that encompasses all of the variables (beyond just a neuron's synaptic inputs) that influence whether a neuron fires or not in a given moment (e.g., delays in integrating inputs, random fluctuations from the release of neurotransmitters in vesicles, firing of variable strength).  As \textcolor{red}{$\beta \rightarrow 0$ ($T \rightarrow \infty$)}, the interactions are governed entirely by randomness.  On the other hand, as \textcolor{red}{$\beta \rightarrow \infty$ ($T \rightarrow 0$)}, the nodal inputs takeover as the only factor in determining the subsequent states of the units---the network becomes deterministic rather than stochastic.

This sigmoidal update rule is commonly used as the nonlinearity in the nodal activation function in stochastic neural networks for reasons coming from statistical mechanics: It arises as a direct consequence of the Boltzmann--Gibbs distribution when assuming pairwise interactions \textcolor{red}{(similar to Glauber dynamics on the Ising model)}, as explained in, for example, \cite{hkp} (Chapter~2 \& Appendix~A).  As a consequence of this update rule, for finite $\beta$ there is always a unique stationary distribution on the stochastic network state space.

%\newpage

\section{Results} \label{results}

What follows are plots comparing and contrasting the four introduced complexity measures in their specified settings.  The qualitative trends shown in the plots empirically hold regardless of network size; a 5-node network was used to generate the plots below.

\begin{figure}[!ht]
\centering

\begin{subfigure}[t]{0.49\textwidth}
\includegraphics[width=\textwidth]{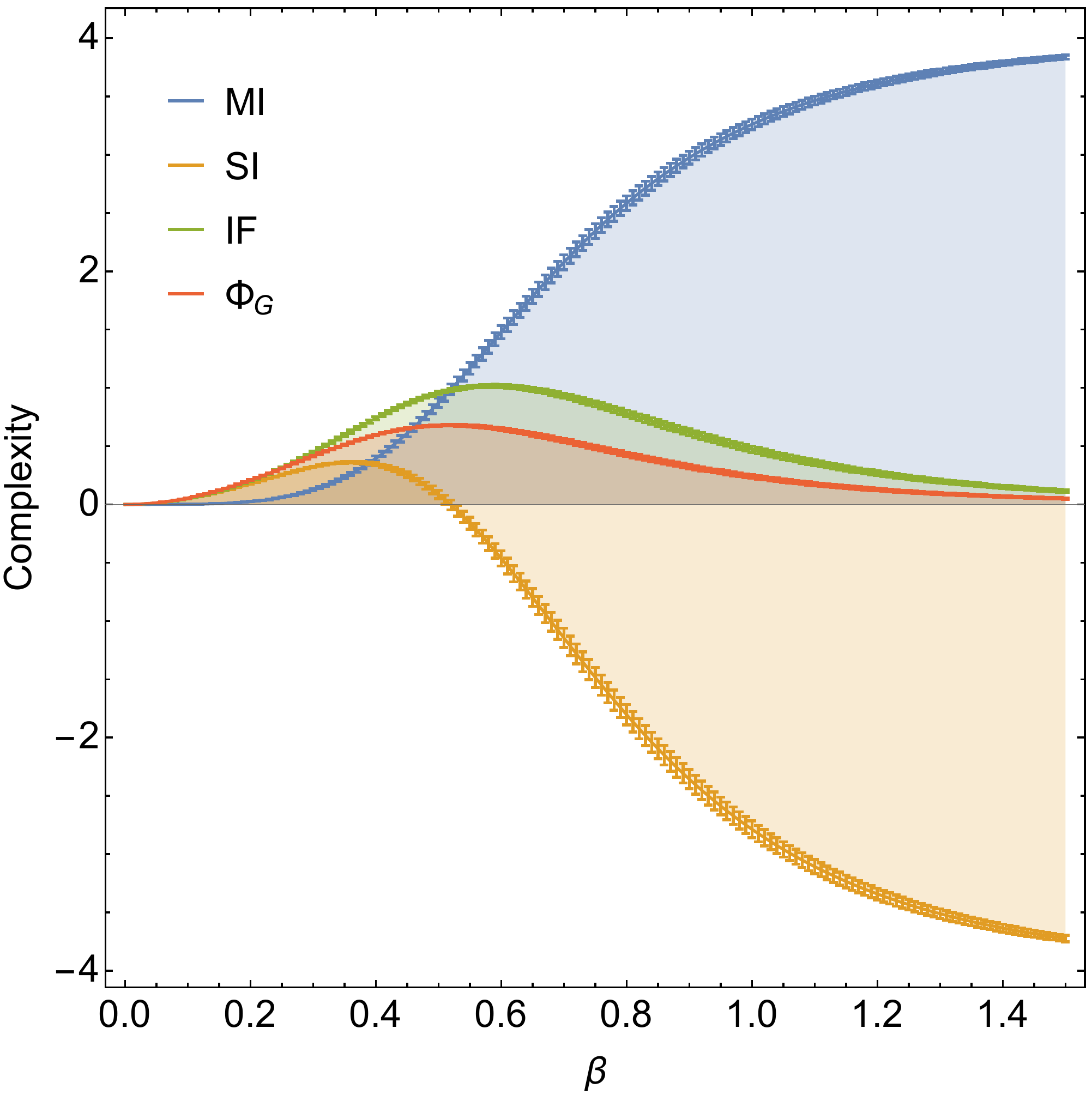}
\caption{Measures of complexity when using random weight initializations sampled uniformly between $0$ and $1$ (averaged over 100 trials, with error bars).}
\label{a}
\end{subfigure}
~
\begin{subfigure}[t]{0.49\textwidth}
\centering
\includegraphics[width=\textwidth]{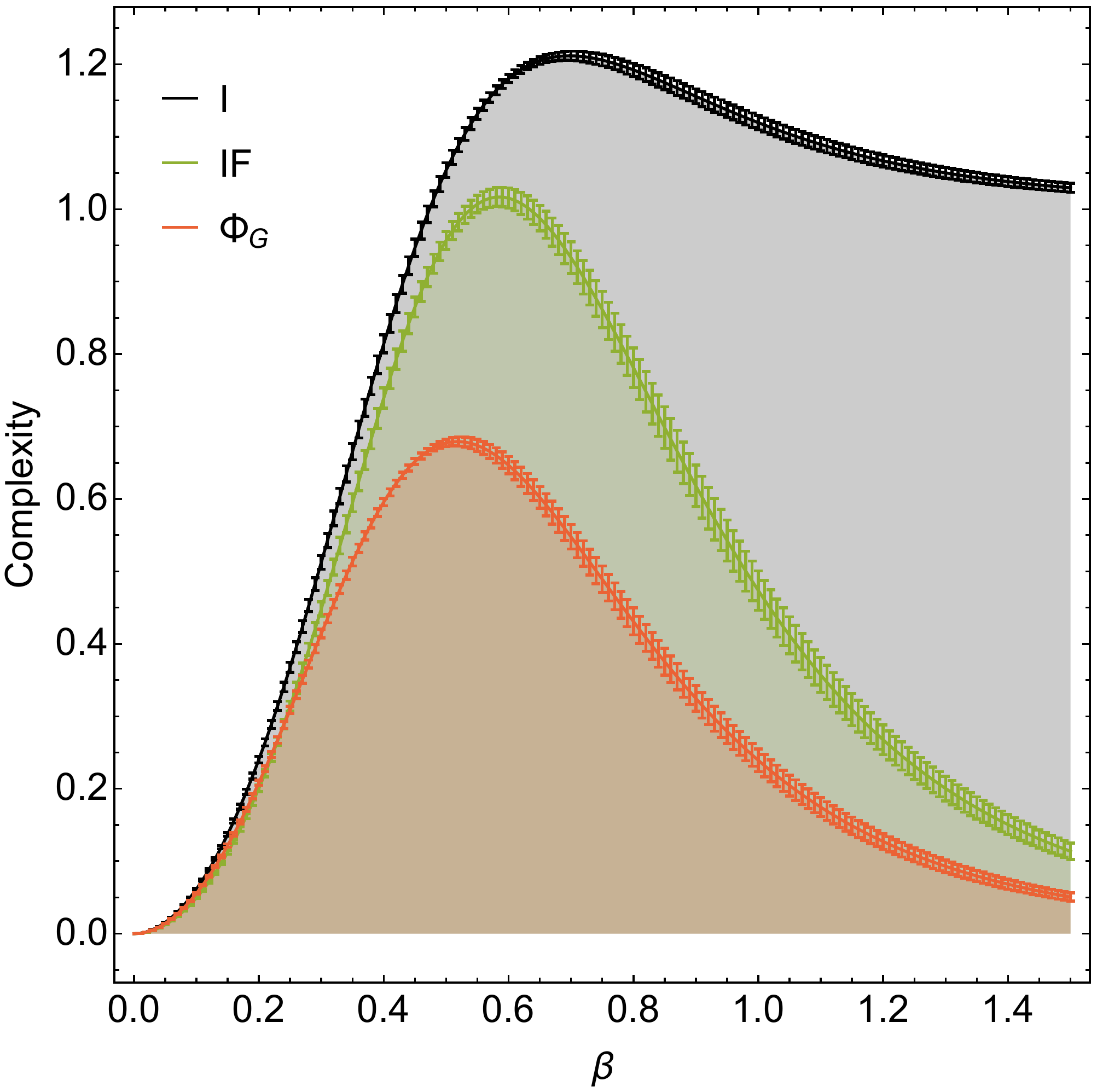}
\caption{\textcolor{red}{The mutual information $I$} upper bounds $IF$ and $\Phi_G$ when using random weight initializations sampled uniformly between $0$ and $1$ (averaged over 100 trials, with error bars).}
\label{b}
\end{subfigure}

%\vspace{-12pt}
\caption{}
\end{figure}

In Figure \ref{a}, we see that when weights are uniformly distributed between $0$ and $1$, $IF$ and $\Phi_G$ are very similar qualitatively, with the additional property that $\Phi_G \leq IF$, which directly follows from ${\mathcal S}^{(1)} \subseteq {\mathcal S}^{(3)}$.  $MI$ monotonically increases, which contradicts the intuition prescribed by humpology.  Finally, $SI$ is peculiar in that it is not lower-bounded by $0$.  This makes for difficult interpretation: What does a negative complexity mean as opposed to zero complexity?  Furthermore, in Figure \ref{b}, we see that $\Phi_G$ satisfies constraint \eqref{postulate}, with the mutual information in fact upper bounding both $IF$ and $\Phi_G$.

It is straightforward to see the symmetry between selecting weights uniformly between $0$ and $+1$ and between $-1$ and $0$, hence the above results represent both scenarios.

\begin{figure}[!ht]
\centering

\begin{subfigure}[t]{0.49\textwidth}
\centering
\includegraphics[width=\textwidth]{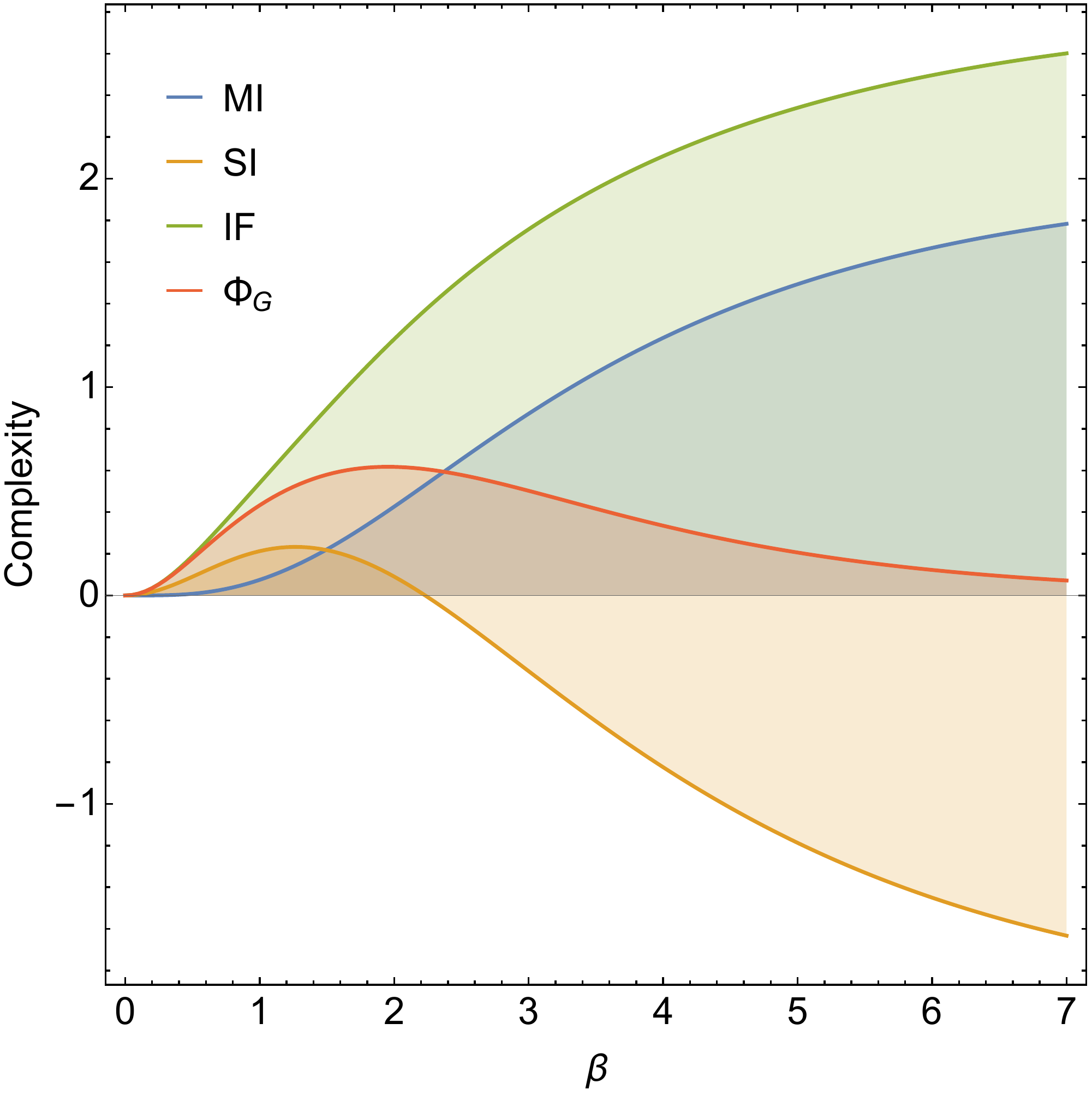}
\caption{\label{c}}
\end{subfigure}
~
\begin{subfigure}[t]{0.49\textwidth}
\centering
\includegraphics[width=\textwidth]{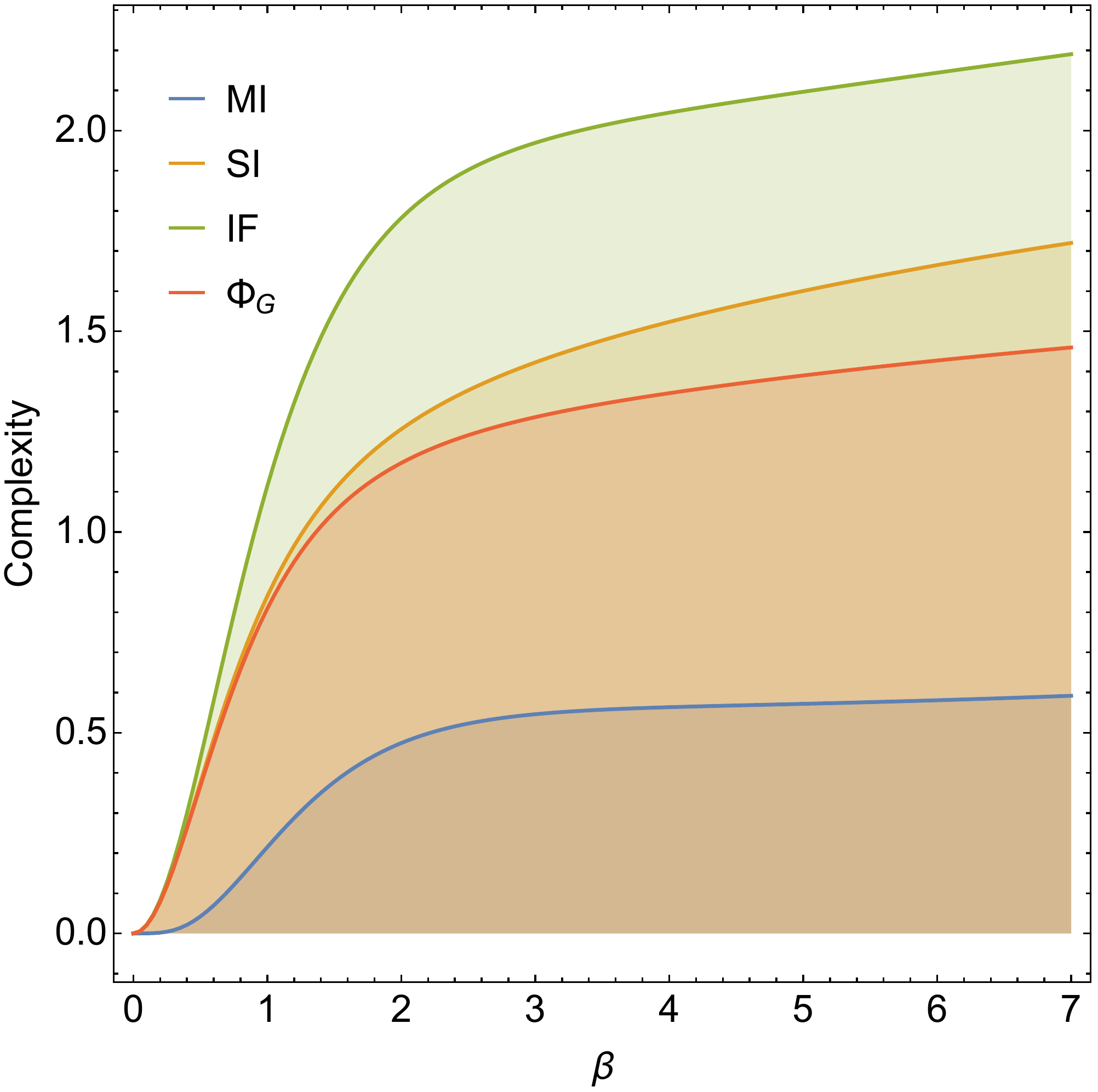}
\caption{\label{d}}
\end{subfigure}
~
\begin{subfigure}[t]{\textwidth}
\centering
\includegraphics[width=0.49\textwidth]{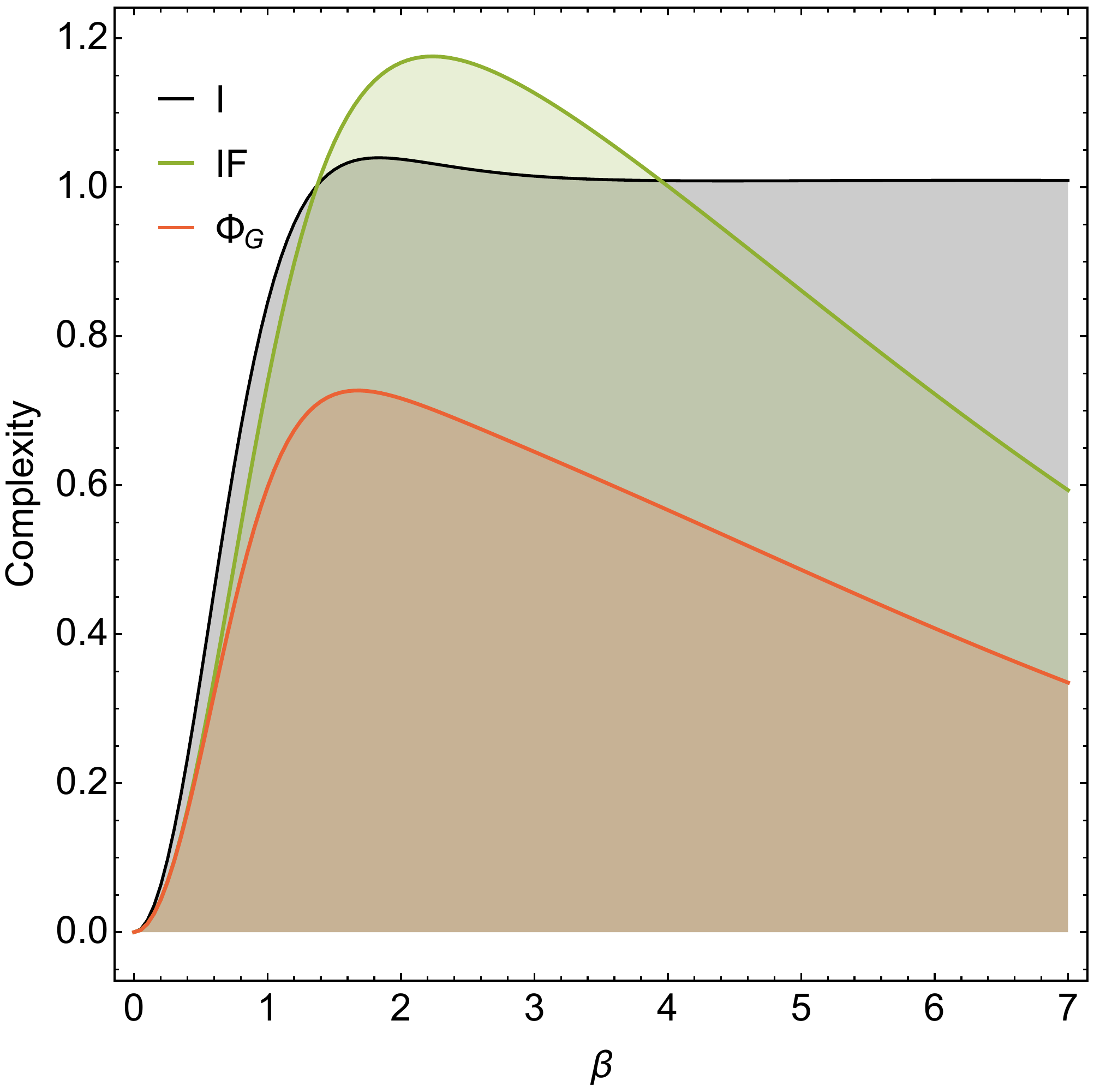}
\caption{\label{e}}
\end{subfigure}

%\vspace{-12pt}
\caption{Measures of complexity in single instances of using random weight initializations sampled uniformly between $-1$ and $1$.}
\label{cde}
\end{figure}

When we allow for both positive and negative weights, however, about as frequently as we observe the above behavior, we observe qualitatively different behavior as represented in Figure~\ref{cde}.  Physically, these results correspond to allowing for mutual excitation and inhibition in the same network.

In Figure \ref{c}, surprisingly, we see that in one instance of mixed weights, $IF$ monotonically increases (like $MI$ in Figure \ref{a}), a departure from the humpology intuition.  Meanwhile, $\Phi_G$ behaves qualitatively differently, such that $\Phi_G \rightarrow 0$ as $\beta \rightarrow \infty$.  In Figure \ref{d}, we see an instance where all measures limit to some non-zero value as $\beta \rightarrow \infty$.  \textcolor{red}{Finally, in Figure \ref{e}, we see an instance where $IF$ exceeds $I$ while $\Phi_G$ satisfies constraint \eqref{postulate}, despite the common unimodality of both measures.}

\textcolor{red}{
An overly simplistic interpretation of the idea that humpology attempts to capture may lead one to believe that Figure \ref{d} is a negative result discrediting all four measures.  We claim, however, that this result suggests that the simple humpology intuition described in Section \ref{intro} needs additional nuance when applied to quantifying the complexity of dynamical systems.  In Figure \ref{d}, we observe a certain richness to the network dynamics, despite its deterministic nature.  A network dynamics that deterministically oscillates around a non-trivial attractor is not analogous to the ``frozen'' state of a rigid crystal (no complexity).  Rather, one may instead associate the crystal state with a network whose dynamics is the identity map, which can indeed be represented by a split stochastic matrix.  Therefore, whenever the stochastic matrix $P$ converges to the identity matrix (the ``frozen'' matrix) for $\beta \to \infty$, the complexity will asymptote to zero (as in Figure \ref{b}).  In other words, for dynamical systems, a ``frozen'' system is exactly that: a network dynamics that has settled into a single \textit{fixed-point} dynamics.  Consequently, in our results, as $\beta \to \infty$, we should expect that the change in complexity depends on the dynamics that the network is settling into as it becomes deterministic, and the corresponding richness (e.g., number of attractors and their lengths) of that asymptotic dynamics.}

\newpage

\begin{figure}[!ht]
\centering
\includegraphics[width=.85\textwidth]{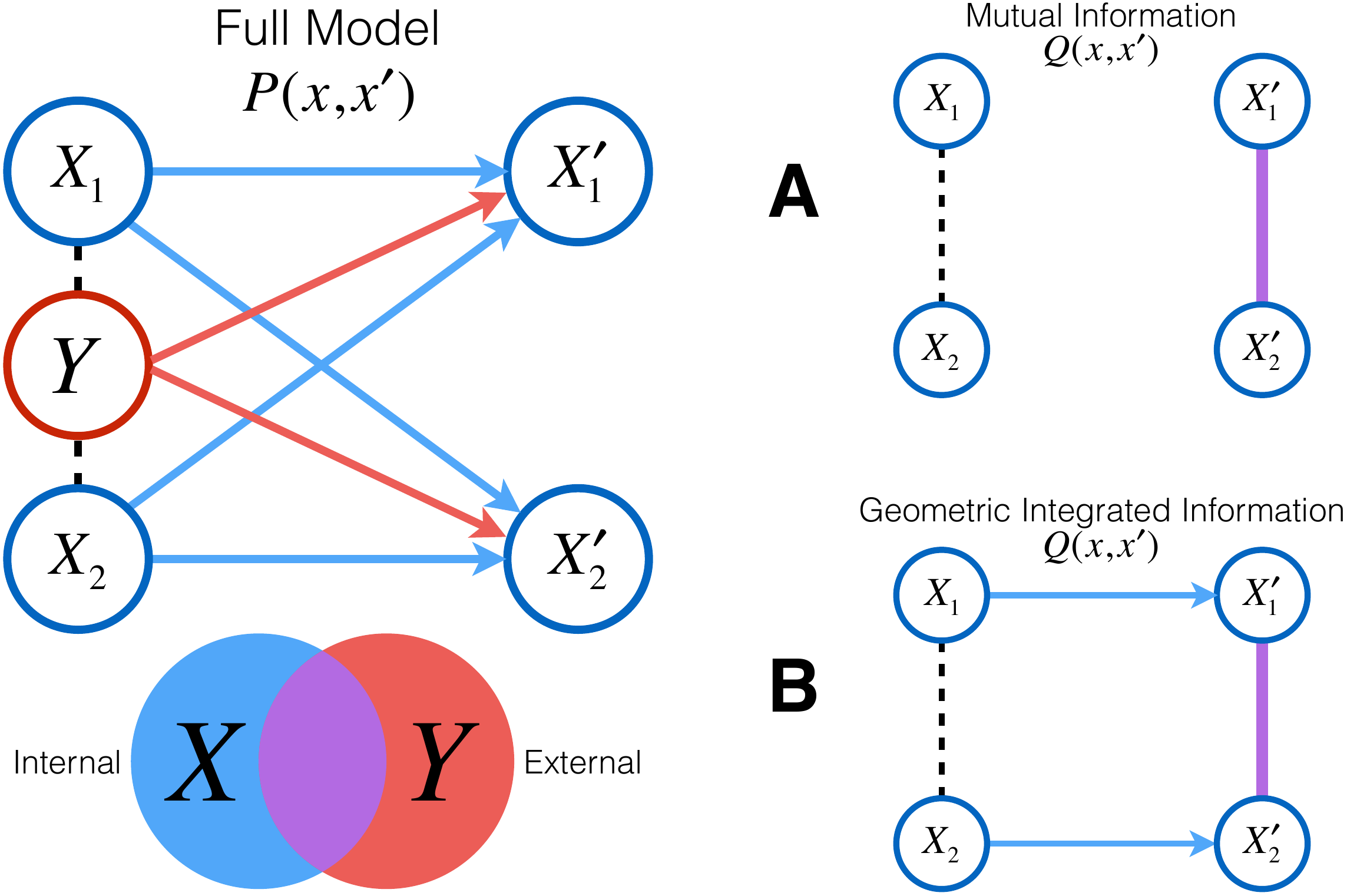}
\caption{\textcolor{red}{A full model (left) can have both intrinsic (blue) and extrinsic (red) causal interactions contributing to its overall dynamics.  Split models (\textbf{A,B}) formulated with an undirected output edge (purple) attempt to exclusively quantify extrinsic causal interactions (so as to strictly preserve intrinsic causal interactions after the ``split''-projection).  However, the output edge can end up explaining away interactions from \textit{both} external factors $Y$ \textit{and} (some) internal factors $X$ (red + blue = purple).  As a result, using such a family of split models does not properly capture the \textit{total} intrinsic causal interactions present in a system.}}
\label{CausalFactors}
\end{figure}

\textcolor{red}{
So far, it may seem to be the case that $\Phi_G$ is without flaw; however, there are shortcomings that warrant further study.  In particular, in formulating $\Phi_G$, the undirected output edge in Figure \ref{CausalFactors}B (purple) was deemed necessary to avoid quantifying external influences to the system that $IF$ would consider as intrinsic information flow.  Yet, in the model studied here---the Boltzmann machine---there are no such external influences (i.e., $Y=0$ in Figure \ref{CausalFactors}), so this modification should have no effect on distinguishing between $\Phi_G$ and $IF$ in our setting.  More precisely, a full model that lacks an undirected output edge at the start should not lead to a ``split''-projection that incorporates such an edge.  However, this is not generally true for the projection that $\Phi_G$ computes because the undirected output edge present in the split model will in fact capture causal interactions \textit{within} the system by deviously interpreting them as same-time interactions in the output (Figure \ref{CausalFactors}).  This counterintuitive phenomenon suggests that we should have preferred $IF$ to be precisely equal to its ideal form $\Phi_{ideal}$ in the case of the Boltzmann machine, and yet, almost paradoxically, this would imply that the improved form would still violate constraint \eqref{postulate}.  This puzzling conundrum begs further study of how to properly disentangle external influences when attempting to strictly quantify the intrinsic causal interactions.}

\textcolor{red}{
The preceding phenomenon, in fact, also calls into question the very postulate that the mutual information ought to be an upper bound on information integration.  As we see in Figure \ref{CausalFactors}A, the undirected output edge used in the ``split''-projection for computing the mutual information $I$ is capable of producing the very same problematic phenomenon.  Thus, the mutual information does not fully quantify the total causal influences intrinsic to a system.  In fact, the assumption itself that $I$ quantified the total intrinsic causal influences was based on the assumption that one can distinguish between intrinsic and extrinsic influences in the first place, which may not be the case.}

% What is information integration $\Phi$? We can use the phenomenon we have observed to in fact define information integration conceptually.  Information integration is all of the intrinsic causal influences that cannot be explained away by an alternative external mechanism.

\section{Application}
\textcolor{red}{In this section, we apply one of the preceding measures ($IF$) and examine its dynamics during network learning.  We wish to exemplify the insights one can gain by exploring measures of complexity in a more general sense.  The results presented in Section \ref{results} showed the promising nature of information-geometric formulations of complexity, such as $IF$ and $\Phi_G$.  Here, however, we restrict ourselves to studying $IF$ as a first step due to the provable properties of its closed-form expression that we are able to exploit to study it in greater depth in the context of autoassociative memory networks.  It would be useful to extend this analysis to $\Phi_G$, but is beyond the scope of this work.}

Autoassociative memory in a network is a form of ``collective computation'' where, given an incomplete input pattern, the network can accurately recall a previously stored pattern by evolving from the input to the stored pattern.  For example, a pattern might be a binary image, in which each pixel in the image corresponds to a node in the network with a value in $\{-1,+1\}$.  In this case, an autoassociative memory model with a stored image could then take as input a noisy version of the stored image and accurately recall the fully denoised original image.  This differs from a ``serial computation'' approach to the same problem where one would simply store the patterns in a database and, when given an input, search all images in the database for the most similar stored image to output.

One mechanism by which a network can achieve collective computation has deep connections to concepts from statistical mechanics (e.g., the Ising model, Glauber dynamics, Gibbs sampling).  This theory is explained in detail in \cite{hkp}.  The clever idea behind autoassociative memory models heavily leverages the existence of an energy function (sometimes called a Lyapunov function) to govern the evolution of the network towards a locally minimal energy state.  Thus, by engineering the network's weighted edges such that local minima in the energy function correspond to stored patterns, one can show that if an input state is close enough (in Hamming distance) to a desired stored state, then the network will evolve towards the correct lower-energy state, which will in fact be a stable fixed point of the network.  
	
The above, however, is only true up to a limit.  A network can only store so many patterns before it becomes saturated.  As more and more patterns are stored, various problems arise such as desirable fixed points becoming unstable optima, as well as the emergence of unwanted fixed points in the network that do not correspond to any stored patterns (i.e., spin glass states).

In 1982, Hopfield put many of these ideas together to formalize what is today known as the Hopfield model, a fully recurrent neural network capable of autoassociative memory.  Hopfield's biggest contribution in his seminal paper was assigning an energy function to the network model:
\begin{equation}
E = -\frac{1}{2}\sum\limits_{i,j} w_{ij}X_i X_j.
\end{equation}

For our study, we assume that we are storing random patterns in the network.  In this scenario, Hebb's rule (Equation \eqref{Hebb}) is a natural choice for assigning weights to each connection between nodes in the network such that the random patterns are close to stable local minimizers of the energy function.

Let $\{\xi^{(1)}, \xi^{(2)},\dots,\xi^{(T)}\}$ denote the set of $N$-bit binary patterns that we desire to store.  Then, under Hebb's rule, the weight between nodes $i$ and $j$ should be assigned as follows:
\begin{equation}\label{Hebb}
w_{ij} = \frac{1}{T} \sum\limits_{\mu = 1}^T \xi^{(\mu)}_i \xi^{(\mu)}_j,
\end{equation}
where $\xi^{(\mu)}_i$ denotes the $i${th}-bit of pattern $\xi^{(\mu)}$.  Notice that all weights are symmetric, $w_{ij} = w_{ji}$.

Hebb's rule is frequently used to model learning, as it is both \textit{local} and \textit{incremental}---two desirable properties of a biologically plausible learning rule.  Hebb's rule is local because weights are set based strictly on local information (i.e., the two nodes that the weight connects) and is incremental because new patterns can be learned one at a time without having to reconsider information from already learned patterns.  Hence, under Hebb's rule, training a Hopfield network is relatively simple and straightforward.  

The update rule that governs the network's dynamics is the same sigmoidal function used in the Boltzmann machine described in Section \ref{rnn}.  We will have this update rule take effect synchronously for all nodes (Note: Hopfield's original model was described in the asynchronous, deterministic case but can also be studied more generally.):
\begin{equation}
\operatorname{Pr}(X^{\prime}_i = +1 \mid X) = \frac{1}{1+e^{-2\beta\sum\limits_{j\in V} X_j \cdot w_{ji}}}.
\end{equation}
\textcolor{red}{At finite $\beta$, our Hopfield model obeys a stochastic sigmoidal update rule.  Thus, there exists a unique and strictly positive stationary distribution of the network dynamics.}

Here, we study \textit{incremental} Hebbian learning, in which multiple patterns are stored in a Hopfield network in succession. We use total information flow (Section \ref{total info flow}) to explore how incremental Hebbian learning changes complexity, or more specifically, how the complexity relates to the number of patterns stored.

Before continuing, we wish to make clear upfront an important disclaimer: The results we describe are qualitatively different when one uses asynchronous dynamics instead of synchronous, as we use here.  With asynchronous dynamics, no significant overall trend manifests, but other phenomena emerge in need of further exploration.

\begin{figure}[!ht]
\centering

\begin{subfigure}[t]{0.49\textwidth}
\centering
\includegraphics[width=\textwidth]{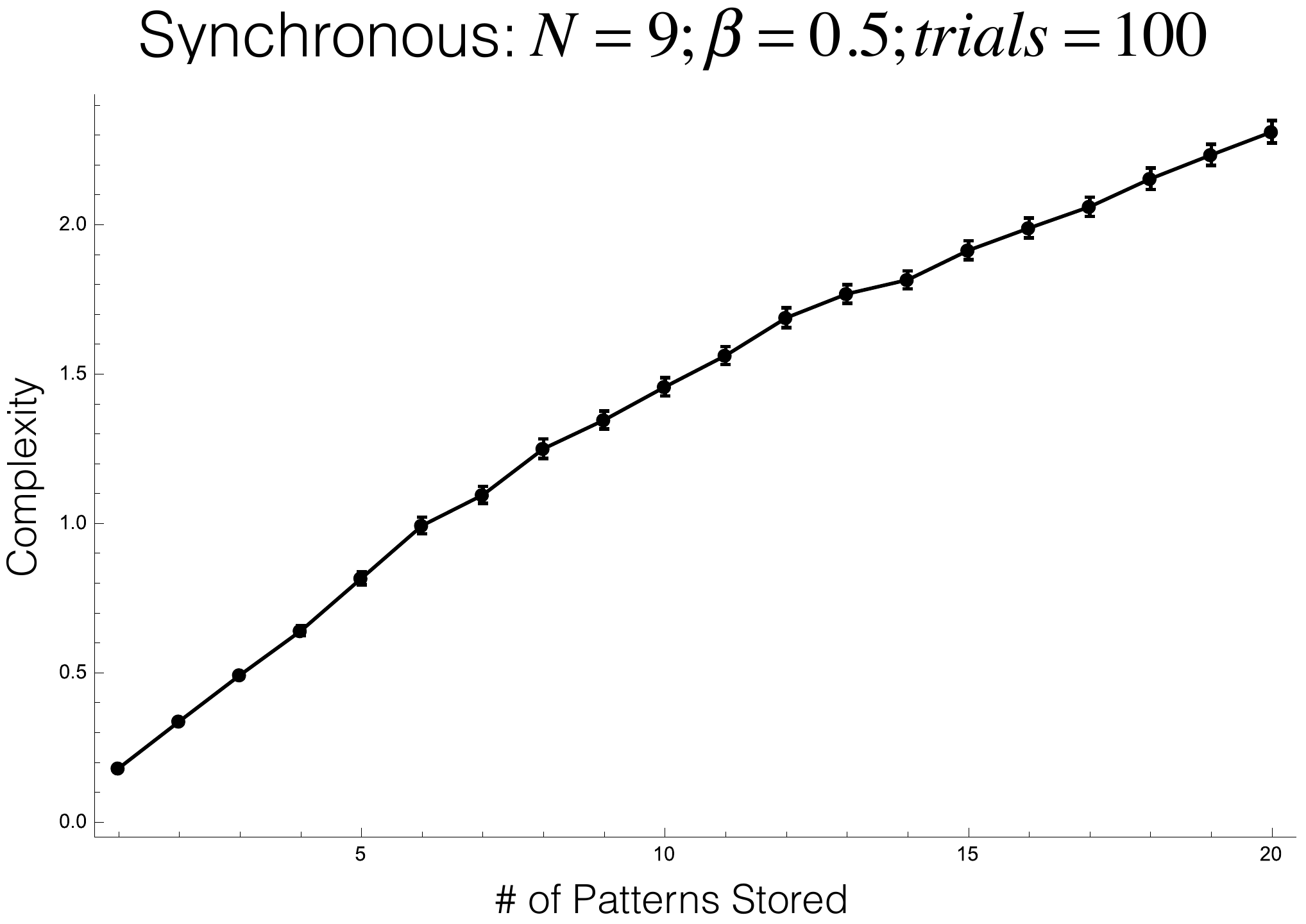}
\caption{$\beta=\frac{1}{2}$}
\label{fig1}
\end{subfigure}
~
\begin{subfigure}[t]{0.49\textwidth}
\centering
\includegraphics[width=\textwidth]{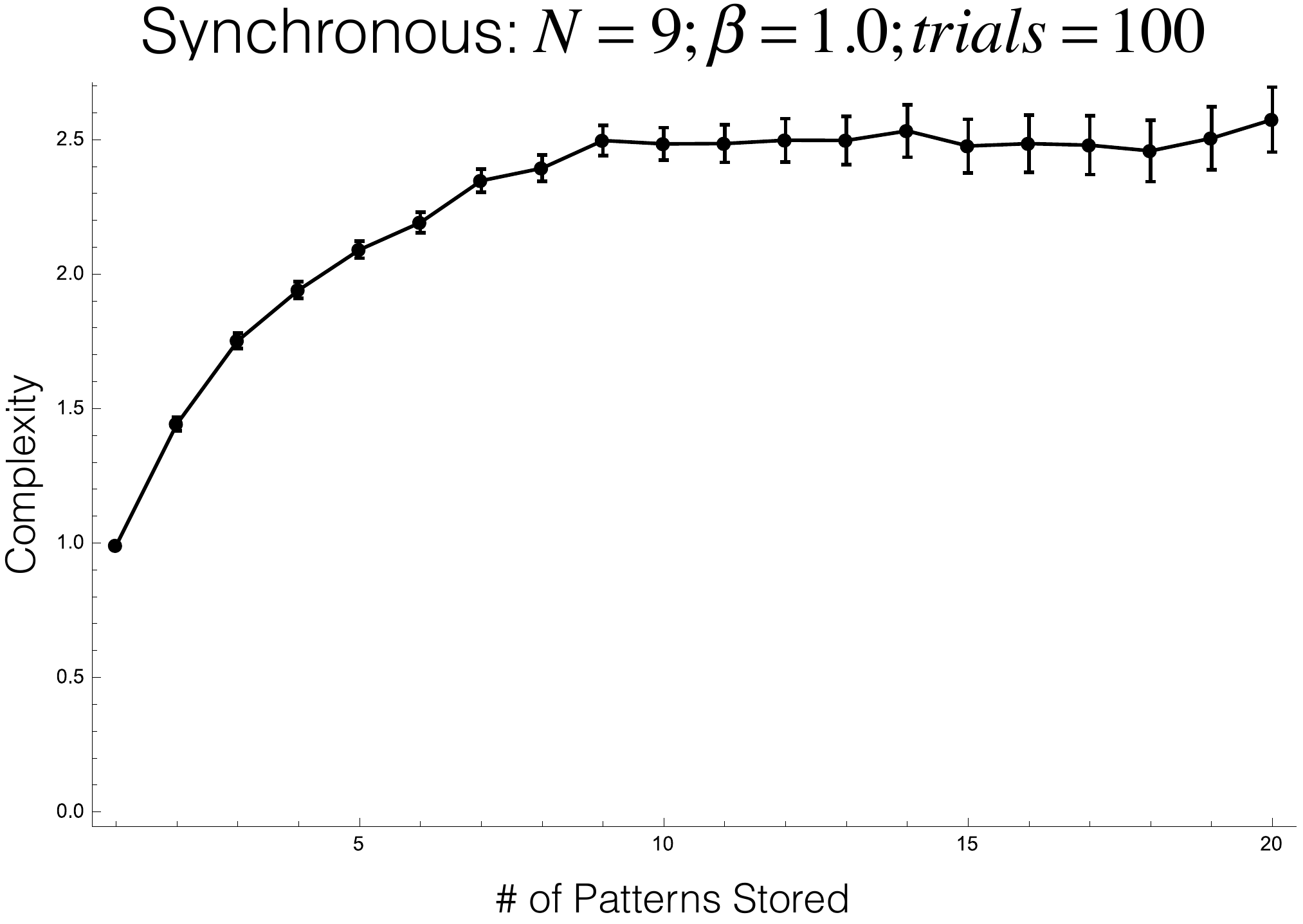}
\caption{$\beta=1$}
\label{fig2}
\end{subfigure}

%\vspace{-12pt}
\caption{Incremental Hebbian learning in a 9-node stochastic Hopfield network with synchronous updating (averaged over 100 trials of storing random 9-bit patterns).}
\end{figure}

When we synchronously update nodes, we see very interesting behavior during learning: incremental Hebbian learning appears to increase complexity, on average (Figures \ref{fig1}, \ref{fig2}).  The dependence on $\beta$ is not entirely clear, but as one can infer from Figures \ref{fig1} and \ref{fig2}, it appears that increasing $\beta$ increases the magnitude of the average complexity while learning, while also increasing the variance of the complexity.  So as $\beta$ increases, the average case becomes more and more unrepresentative of the individual cases of incremental Hebbian learning.

We can also study the deterministic version of the Hopfield model.  This corresponds to letting $\beta \rightarrow \infty$ in the stochastic model.  With a deterministic network, many stationary distributions on the network dynamics may exist, unlike \textcolor{red}{in} the stochastic case.  As discussed above, if we want to recall a stored image, we would like for that image to be a fixed point in the network (corresponding to a stationary distribution equal to the Dirac measure at that state).  Storing multiple images corresponds to the desire to have multiple Dirac measures acting as stationary distributions of the network.  Furthermore, in the deterministic setting the nodal update rule becomes a step rather than a sigmoid function.

Without a unique stationary distribution in the deterministic setting, we must decide how to select an input distribution to use in calculating the complexity.  If there are multiple stationary distributions in a network, not all starting distributions on the network eventually lead to a single stationary distribution (as was the case in the stochastic model), but instead the stationary distribution that the network eventually reaches is sensitive to the initial state of the network.  When there are multiple stationary distributions, there are actually infinitely many stationary distributions, as any convex combination of stationary distributions is also stationary.  If there exist $N$ orthogonal stationary distributions of a network, then there is in fact an entire $(N-1)$-simplex of stationary distributions, any of which could be used as the input distribution for calculating the complexity.

In order to address this issue, it is fruitful to realize that the complexity measure we are working with is concave with respect to the input distribution (Theorem \ref{concavity} in \hyperlink{appendix}{Appendix A}).  As a function of the input distribution, there is thus an ``apex'' to the complexity.  In other words, is a unique local maximum of the complexity function, which is also therefore a global maximum (but not necessarily a unique maximizer since the complexity is not strictly concave).  This means that the optimization problem of finding the supremum over the entire complexity landscape with respect to the input distribution is relatively simple and can be viably achieved via standard gradient-based methods.

We can naturally define a new quantity to measure complexity of a stochastic matrix $P$ in this setting, the \textit{complexity capacity}:
\begin{equation}\label{capacity}
C_{cap}(X\rightarrow X' \mid P) \triangleq \max_{p} C(X\rightarrow X' \mid p, P),
\end{equation}
where the maximum is taken over all stationary distributions $p$ of $P$.   Physically, the complexity capacity measures the \textit{maximal} extent---over possible input distributions---to which the whole is more than the sum of its parts.  By considering the entire convex hull of stationary input distributions and optimizing for complexity, we can find this unique maximal value and use it to represent the complexity of a network with multiple stationary distributions.

\begin{figure}[!ht]
\centering

\begin{subfigure}[t]{0.49\textwidth}
\centering
\includegraphics[width=\textwidth]{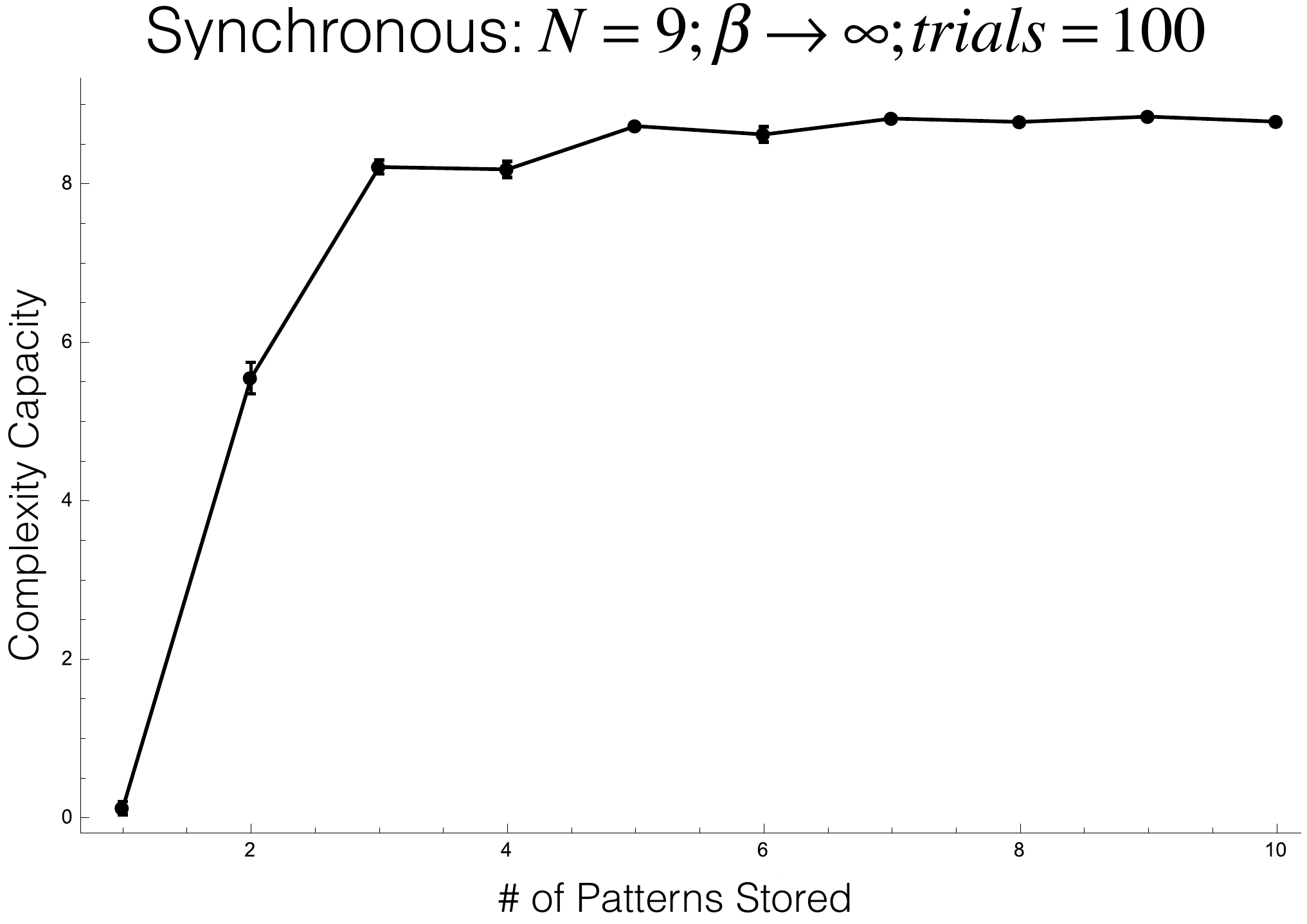}
\caption{$N=9$}
\label{fig3}
\end{subfigure}
~
\begin{subfigure}[t]{0.49\textwidth}
\centering
\includegraphics[width=\textwidth]{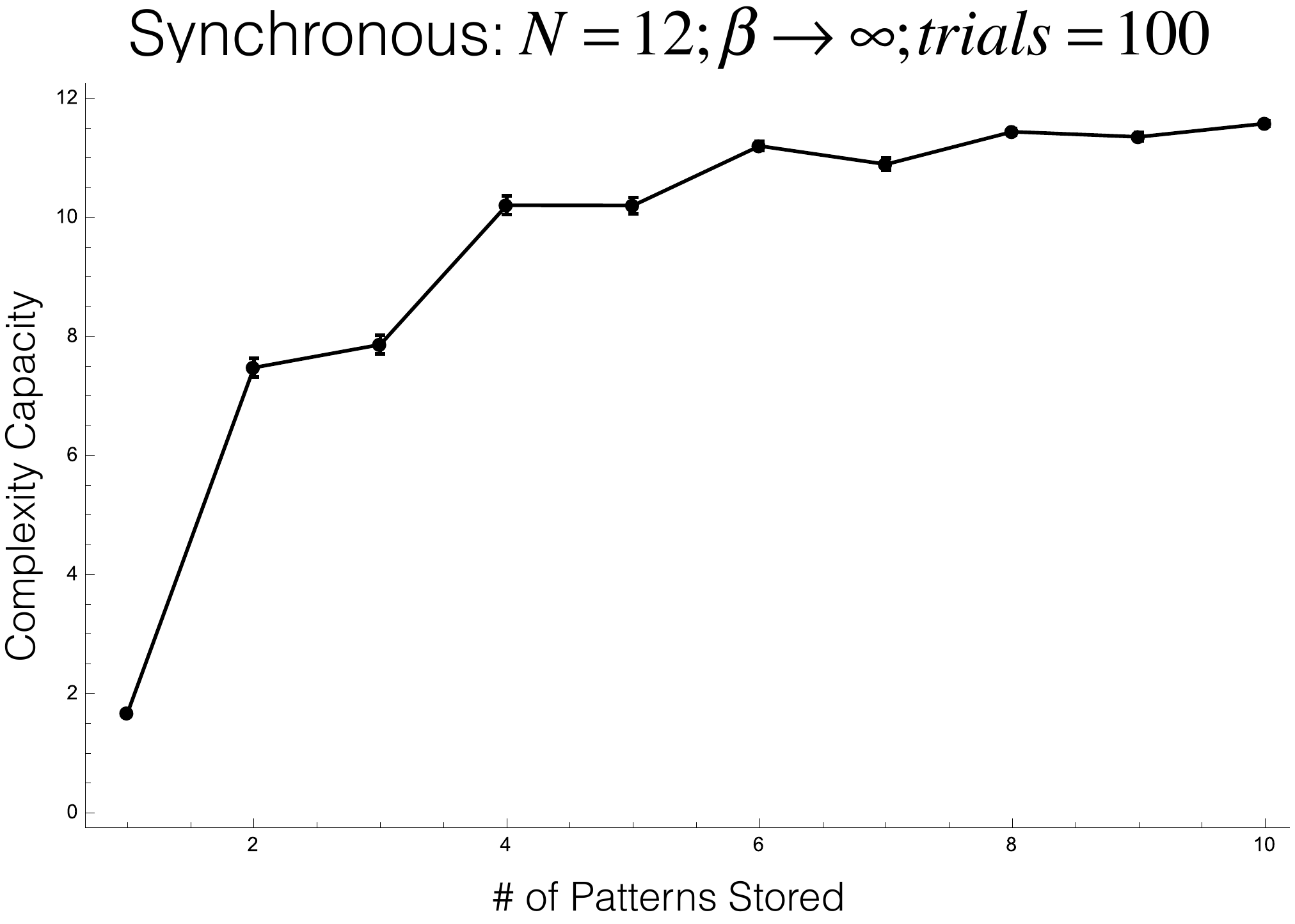}
\caption{$N=12$}
\label{fig4}
\end{subfigure}

%\vspace{-12pt}
\caption{Incremental Hebbian learning in a $N$-node deterministic ($\beta\rightarrow \infty$) Hopfield network with synchronous updating (averaged over 100 trials of storing random $N$-bit patterns).}
\end{figure}

Again, in the synchronous-update setting, we see incremental Hebbian learning increases complexity \textit{capacity} (Figures \ref{fig3}, \ref{fig4}).  It is also worth noting that the complexity capacity in this setting is limiting towards the absolute upper bound on the complexity, which can never exceed the number of binary nodes in the network.  Physically, this corresponds to each node attempting to store one full bit (the most information a binary node can store), and all of this information flowing through the network between time-steps, as more and more patterns are learned.  This limiting behavior of the complexity capacity towards a maximum (as the network saturates with information) is more gradual as the size of the network increases.  This observed behavior matches the intuition that larger networks should be able to store more information than smaller networks.

\section{Conclusions}
In summary, we have seen four different measures of complexity applied in concrete, parameterized systems.  We observed that the synergistic information was difficult to interpret on its own due to the lack of an intuitive lower bound on the measure.  Building off the primitive multi-information, the total information flow and the geometric integrated information were closely related, frequently (but not always) showing the same qualitative behavior.  The geometric integrated information satisfies the additional postulate \eqref{postulate} stating that a measure of complexity should not exceed the temporal mutual information, a property that the total information flow frequently violated in the numerical experiments where connection weights were allowed to be both negative and positive.  The geometric integrated information was recently proposed to build \textcolor{red}{on} and correct the original flaws in the total information flow, which it appears to have done quite singularly based on the examination in the present study.  \textcolor{red}{While the geometric integrated information is a step in the right direction, further study is needed to properly disentangle external from internal causal influences that contribute to network dynamics (see final paragraphs of Section \ref{results}).  Nonetheless,} it is encouraging to see a semblance  of convergence with regards to quantifying complexity from an information-theoretic perspective.

%%%%%%%%%%%%%%%%%%%%%%%%%%%%%%%%%%%%%%%%%%
\vspace{6pt} 

%%%%%%%%%%%%%%%%%%%%%%%%%%%%%%%%%%%%%%%%%%
\acknowledgments{The authors would like to thank the Santa Fe Institute NSF REU program (NSF grant \#1358567), where this work began, and the Max Planck Institute for Mathematics in the Sciences (DFG Priority Program 1527 Autonomous Learning), where this work continued.  J.A.G. was supported by an SFI Omidyar Fellowship during this work.}

%%%%%%%%%%%%%%%%%%%%%%%%%%%%%%%%%%%%%%%%%%
\authorcontributions{N.A. and J.A.G. proposed the research. M.S.K. carried out most of the research and took the main responsibility for writing the article. All authors contributed to joint discussions. All authors read and approved the final manuscript.}

%%%%%%%%%%%%%%%%%%%%%%%%%%%%%%%%%%%%%%%%%%
\conflictsofinterest{The authors declare no conflict of interest.} 

%%%%%%%%%%%%%%%%%%%%%%%%%%%%%%%%%%%%%%%%%%
%% optional
\appendixtitles{no} %Leave argument "no" if all appendix headings stay EMPTY (then no dot is printed after "Appendix A"). If the appendix sections contain a heading then change the argument to "yes".
\appendixsections{multiple} %Leave argument "multiple" if there are multiple sections. Then a counter is printed ("Appendix A"). If there is only one appendix section then change the argument to "one" and no counter is printed ("Appendix").
\hypertarget{appendix}
\appendix
\section{}
\begin{Theorem}[Concavity of $IF(X\rightarrow X')$] \thlabel{concavity}
The complexity measure
\[IF(X\rightarrow X') \triangleq \sum\limits_{v \in V} H(X^{\prime}_v \mid X_v) - H(X' \mid X),\]
is concave with respect to the input distribution $p(x) = \operatorname{Pr}(X = x)$, $x \in \Bbb{X}$, for stochastic matrix $P$ fixed.
\end{Theorem}

\textcolor{red}{Note that in the definition of the \textit{complexity capacity} \eqref{capacity}, we take the supremum over all \textit{stationary} input distributions.  Since such distributions form a convex subset of the set of all input distributions, concavity of $IF$ is preserved by the corresponding restriction.}

\begin{proof}
The proof of the above statement follows from first rewriting the complexity measure in terms of a negative KL divergence between two distributions both affine with respect to the input distribution, and then using the fact that the KL divergence is convex with respect to a pair of distributions (see~\cite{Cover&Thomas}~(Chapter~2)) to demonstrate that the complexity measure is indeed concave.

Let $P$ denote the fixed stochastic matrix governing the evolution of $X \rightarrow X'$.

Let $p$ denote the input distribution on the states of $X$.

First, note that the domain of $p$ forms a convex set: For an $N$-unit network, the set of all valid distributions $p$ \textcolor{red}{forms} an $(N-1)$-simplex.

Next, we expand $IF$:
\vspace{-1.5pt}
\begin{eqnarray*}
IF(X\rightarrow X') & = &\sum\limits_{v \in V} H(X^{\prime}_v \mid X_v) - H(X' \mid X)\\
& = &-\sum\limits_{v \in V} \left(\sum\limits_{x_v \in \Bbb{X}_v} \operatorname{Pr}(X_v = x_v) \sum\limits_{{x'_v \in \Bbb{X}_v}} \operatorname{Pr}(X'_v = x'_v \mid X_v = x_v) \cdot \log{\operatorname{Pr}(X'_v = x'_v \mid X_v = x_v)}\right) \\ && + \sum\limits_{x \in \Bbb{X}} \operatorname{Pr}(X=x)\sum\limits_{x' \in \Bbb{X}} \operatorname{Pr}(X'=x' \mid X=x)\cdot \log{ \operatorname{Pr}(X'=x' \mid X=x)}.
\end{eqnarray*}
Notice that the expanded expression for $H(X' \mid X)$ is affine in the input distribution $p$, since the terms $\operatorname{Pr}(X'=x' \mid X=x)$ are just constants given by $P(x,x')$.  Hence, $-H(X' \mid X)$ is concave, and all that is left to show is that the expansion of $H(X^{\prime}_v \mid X_v)$ is also concave for all $v \in V$:
\begin{eqnarray*}
H(X^{\prime}_v \mid X_v) &=& -\sum\limits_{x_v \in \Bbb{X}_v} \operatorname{Pr}(X_v = x_v) \sum\limits_{{x'_v \in \Bbb{X}_v}} \operatorname{Pr}(X'_v = x'_v \mid X_v = x_v) \cdot \log{\operatorname{Pr}(X'_v = x'_v \mid X_v = x_v)}\\
&=& -\sum\limits_{x_v \in \Bbb{X}_v}\sum\limits_{{x'_v \in \Bbb{X}_v}} \operatorname{Pr}(X'_v = x'_v, X_v = x_v) \cdot \log{\frac{\operatorname{Pr}(X'_v = x'_v, X_v = x_v)}{\operatorname{Pr}(X_v = x_v)}}\\
&=& -\sum\limits_{x_v \in \Bbb{X}_v}\sum\limits_{{x'_v \in \Bbb{X}_v}} \operatorname{Pr}(X'_v = x'_v, X_v = x_v) \cdot \log{\frac{\operatorname{Pr}(X'_v = x'_v, X_v = x_v)}{\operatorname{Pr}(X_v = x_v)}} \\ && + \log{\frac{1}{|\Bbb{X}_v|}} - \log{\frac{1}{|\Bbb{X}_v|}}\\
&=& -\sum\limits_{x_v \in \Bbb{X}_v}\sum\limits_{{x'_v \in \Bbb{X}_v}} \operatorname{Pr}(X'_v = x'_v, X_v = x_v) \cdot \log{\frac{\operatorname{Pr}(X'_v = x'_v, X_v = x_v)}{\operatorname{Pr}(X_v = x_v)}} \\ && +\sum\limits_{x_v \in \Bbb{X}_v}\sum\limits_{{x'_v \in \Bbb{X}_v}} \operatorname{Pr}(X'_v = x'_v, X_v = x_v)\log{\frac{1}{|\Bbb{X}_v|}} \\&& - \log{\frac{1}{|\Bbb{X}_v|}}\\
&=& -\mathlarger{\sum\limits_{x_v \in \Bbb{X}_v}\sum\limits_{{x'_v \in \Bbb{X}_v}}} \bigg( \operatorname{Pr}(X'_v = x'_v, X_v = x_v) \cdot \log{\frac{\operatorname{Pr}(X'_v = x'_v, X_v = x_v)}{\frac{1}{|\Bbb{X}_v|}\cdot \operatorname{Pr}(X_v = x_v)}}\bigg) - \log{\frac{1}{|\Bbb{X}_v|}}.
\end{eqnarray*}
Ignoring the constant $-\log{\frac{1}{|\Bbb{X}_v|}}$, as this does not change the concavity of the expression, we can rewrite the summation as
\begin{eqnarray*}
= -\mathlarger{\sum\limits_{x_v \in \Bbb{X}_v}\sum\limits_{{x'_v \in \Bbb{X}_v}}} \Bigg( & \bigg( \mathlarger{\sum\limits_{x_r \in \Bbb{X}_{V\backslash v}}} \operatorname{Pr}(X'_v = x'_v \mid X_v = x_v, X_{V\backslash v}= x_r)\cdot \operatorname{Pr}(X_v = x_v, X_{V\backslash v}= x_r) \bigg) \cdot &\\
&\log{\frac{\mathlarger{\sum_{x_r \in \Bbb{X}_{V\backslash v}}} \operatorname{Pr}(X'_v = x'_v \mid X_v = x_v, X_{V\backslash v}= x_r)\cdot \operatorname{Pr}(X_v = x_v, X_{V\backslash v}= x_r)}{\frac{1}{|\Bbb{X}_v|}\cdot \mathlarger{\sum_{x_r \in \Bbb{X}_{V\backslash v}}} \operatorname{Pr}(X_v = x_v, X_{V\backslash v} = x_r)}} &\Bigg),
\end{eqnarray*}
where $X_{V\backslash v}$ denotes the state of all nodes excluding $X_v$.  This expansion has made use of the fact that $\operatorname{Pr}(X'_v = x'_v, X_v = x_v) = \sum_{x_r \in \Bbb{X}_{V\backslash v}}\operatorname{Pr}(X'_v = x'_v \mid X_v = x_v, X_{V\backslash v} = x_r) \cdot \operatorname{Pr}(X_v = x_v, X_{V\backslash v}=x_r)$ and $\operatorname{Pr}(X_v = x_v) = \sum_{x_r \in \Bbb{X}_{V\backslash v}} \operatorname{Pr}(X_v = x_v, X_{V\backslash v} = x_r)$.  

The constant $\operatorname{Pr}(X'_v = x'_v \mid X_v = x_v, X_{V\backslash v}= x_r) = \operatorname{Pr}(X'_v = x'_v \mid X = (x_v, x_r))$ can be computed directly as a marginal over the stochastic matrix $P$.  Furthermore, the constant $\operatorname{Pr}(X_v = x_v, X_{V\backslash v}= x_r) = \operatorname{Pr}(X=(x_v,x_r))$ comes directly from the input distribution $p$, making the entire expression for $\operatorname{Pr}(X'_v = x'_v, X_v = x_v)$ affine with respect to the input distribution.

Finally, we get
\begin{eqnarray*}
&=& - D_{KL}\left(\sum\limits_{x_r \in \Bbb{X}_{V\backslash v}}\operatorname{Pr}(X'_v = x'_v \mid X_v = x_v, X_{V\backslash v}=x_r)\cdot \operatorname{Pr}(X_v = x_v, X_{V\backslash v}=x_r)\  \mathlarger{\mathlarger{\mathlarger{\|}}} \right.\\
& &\left. \qquad\qquad \frac{1}{|\Bbb{X}_v|} \cdot \sum\limits_{x_r \in \Bbb{X}_{V\backslash v}} \operatorname{Pr}(X_v = x_v, X_{V\backslash v} = x_r)\right) \\
&=& -D_{KL}\left(\operatorname{Pr}(X'_v = x'_v, X_v = x_v)\ \mathlarger{\mathlarger{\mathlarger{\|}}}\ \frac{1}{|\Bbb{X}_v|}\cdot \operatorname{Pr}(X_v = x_v)  \right),
\end{eqnarray*}
the KL divergence between two distributions, both of which have been written so as to explicitly show them as affine in the input distribution $p$, and then simplified to show that both are valid joint distributions over the states on the pair $(X^{\prime}_v, X_v)$.  Thus, the overall expression is concave with respect to the input distribution.
\end{proof}

%%%%%%%%%%%%%%%%%%%%%%%%%%%%%%%%%%%%%%%%%%
%=====================================
% References, variant B: external bibliography
%=====================================
\externalbibliography{yes}
\bibliography{main}

%%%%%%%%%%%%%%%%%%%%%%%%%%%%%%%%%%%%%%%%%%
\end{document}